# Web Usage Analysis: New Science Indicators and Co-usage

Xavier Polanco, Ivana Roche, Dominique Besagni
{polanco,roche,besagni}@inist.fr

Institut de l'Information Scientifique et Technique (INIST / CNRS)
2 allée du Parc de Brabois – 54514 Vandoeuvre-lès-Nancy – France

## Résumé

A new type of statistical analysis of the science and technical information (STI) in the Web context is produced. We propose a set of indicators about Web users, visualized bibliographic records, and e-commercial transactions. In addition, we introduce two Web usage factors. Finally, we give an overview of the co-usage analysis. For these tasks, we introduce a computer based system, called Miri@d, which produces descriptive statistical information about the Web users' searching behaviour, and what is effectively used from a free access digital bibliographical database. The system is conceived as a server of statistical data which are carried out beforehand, and as an interactive server for online statistical work. The results will be made available to analysts, who can use this descriptive statistical information as raw data for their indicator design tasks, and as input for multivariate data analysis, clustering analysis, and mapping. Managers also can exploit the results in order to improve management and decision-making.

## 1 Introduction

Two scientific communities are dealing with Web analysis related questions. This is the reason why we can observe in the literature two traditions about the analysis of the Web. One developed by people coming from documentation, and the other by computer scientists. The first was developed in the field of information science under the appellations of "webometrics" (Almind & Ingwersen, 1997), or "cybermetrics" (cf. http://www.cindoc.csic.es/cybermetrics) while seeking to extend the informetric techniques to the analysis of the Web (Björneborn & Ingwersen, 2001; Ingwersen & Björneborn, 2004). The second one arose in the field of the computer science while seeking to extend the data mining techniques to Web analysis under the appellation of "Web mining" (Chakrabarti, 2003) and according to three main categories: Web structure mining, Web content mining, and Web usage mining (Kosala & Blockeel, 2000). We work at the border of these two traditions: we consider informetrics from the point of view of computer-based technologies. The Web represents a new environment for the quantitative studies of science, and a new family of computer-based science indicators can be developed. This article deals with a system able to produce descriptive bibliometric statistics, and statistical information on Web users' behaviour.

The article is organized as follows. The first two sections deal with the presentation of the Miri@d server (section 2), and the statistical indicators that Miri@d is able to produce (section 3). The results of the Miri@d application are exposed in section 4. Section 5 describes the co-usage analysis and section 6 deals with the application of co-usage analysis on Web user data coming from the Miri@d server.

# 2 Server organisation

We provide in this section a detailed description of the Miri@d server structure. We start distinguishing the conceptual model that Miri@d represents and its actual technological implementation. The first is general and the second is local.

## 2.1 The model

Figure 1 represents what we call the model. The model is general in the sense that it is not limited to the particular characteristics represented in figure 2. It is significant to see that the model implies three families of data which it can exploit on the one hand log-files data and on the other hand bibliographic data and commercial data. From the economic point of view, the bibliographic database can be replaced by the concept of an unspecified product database. From the point of view of scientific information, the bibliographic database can also be any. At least theoretically, i.e., on the level of its concept, the model is not completely enclosed within the data sources which it is today using.

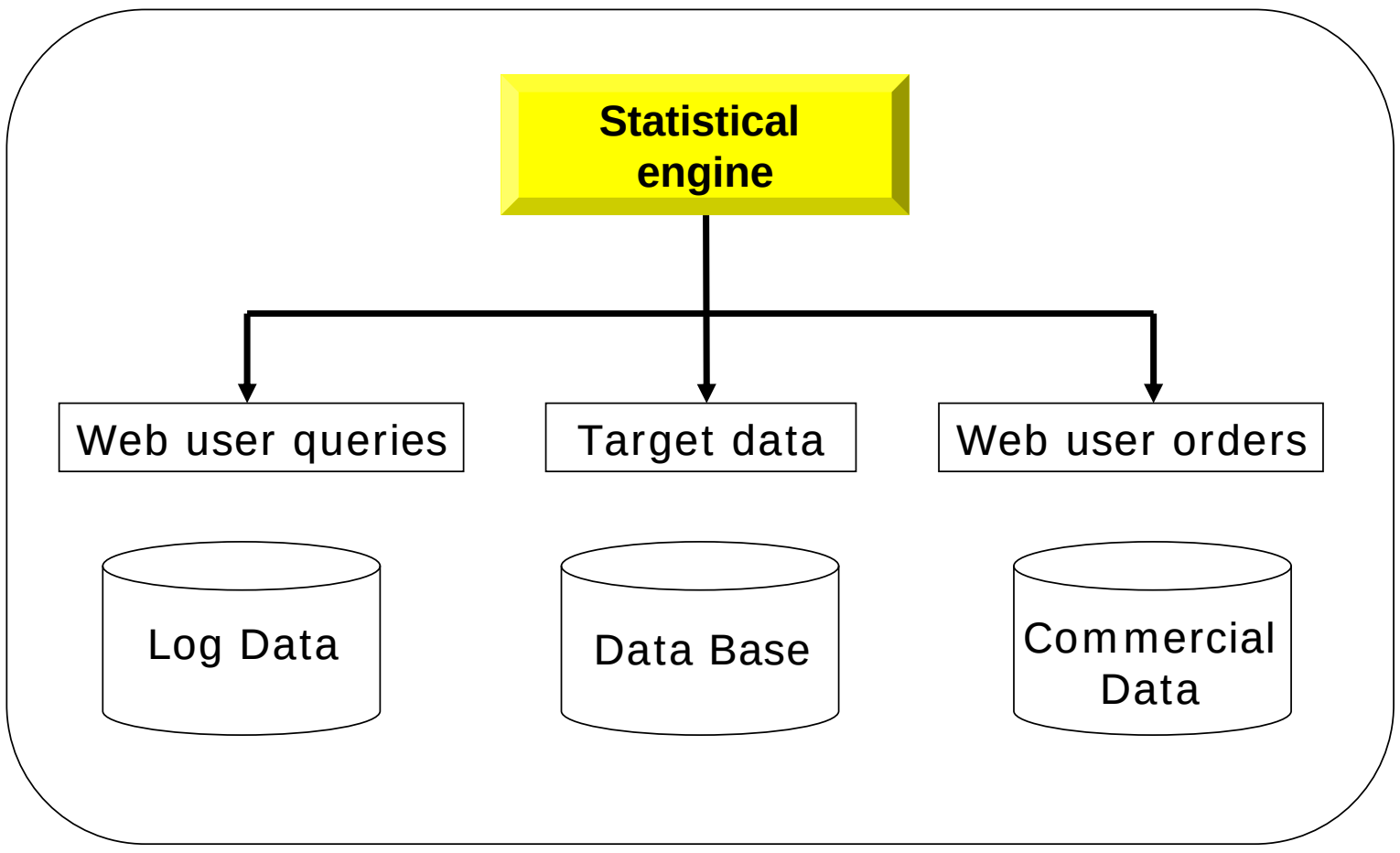

**Figure 1: The model**

## 2.2 The server structure

Figure 2 represents the server structure, which consists of a set of external resources that providing raw data, and a set of database internal to server.

The resources from which Miri@d receives data:

| | |
|---|---|
| DM | document delivery management system |
| CM | customer management system |
| LM | library management system and |
| Article@INIST | server that provides both bibliographic database and log-files |

The Miri@d server own databases:

| | |
|---|---|
| QUERY | data related to Web users queries |
| DISPLAY | data related to displayed bibliographical records |
| ORDER | data related to ordered documents |
| BIBLIO | bibliographical records |
| STAT | calculated statistical indicators |

The data we can obtain from DM and CM systems are processed in order to match information about customers' document orders and customers' characteristics like their geographical location or activity. The collected data from LM permit essentially to complete the characterisation of the bibliographic notices coming from Article@INIST; for example, the principal scientific domains related to each journal concerned by a document order or a bibliographic notice displayed. These data are extracted under request and are used to enrich the records of Miri@d own bibliographic database, denoted by BIBLIO.

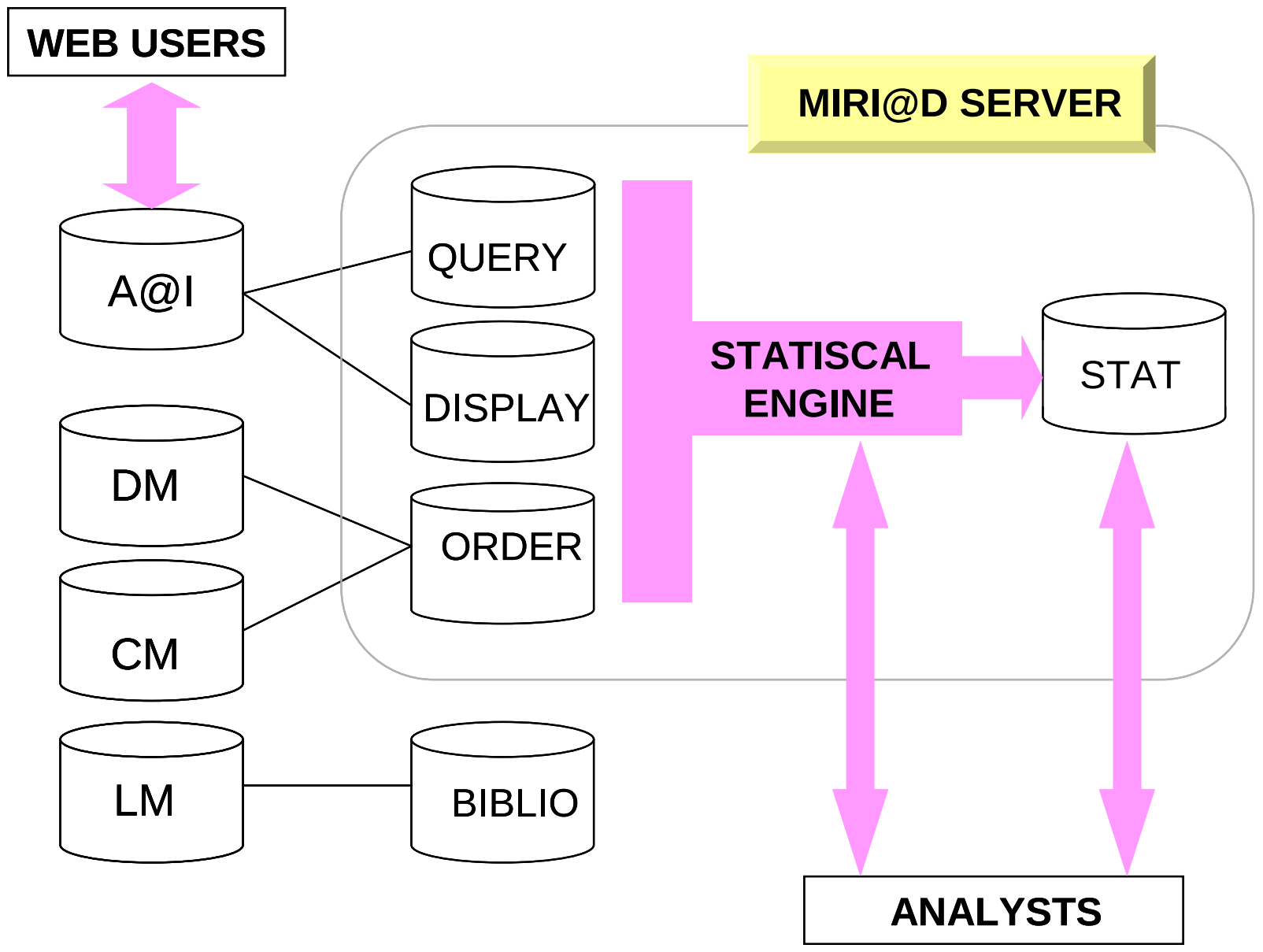

**Figure 2: The server**

The bibliographic data coming from Article@INIST databases are obtained by extraction under request and activated by the identification of each new document ordered or new displayed bibliographic notice. The data obtained from the log-files of Article@INIST provide information about users' behaviour. These data contain information about both users' queries and displayed bibliographic references. The collected data from LM permit essentially to complete the characterisation of the bibliographic notices coming from Article@INIST.

STAT database stores a priori calculated statistics on the data stored in the other Miri@d databases: QUERY, DISPLAY, ORDER and BIBLIO. The user must express his preferences choosing a periodicity, then a period and finally statistical information. The statistics are calculated off-line with a periodicity of one day, one week, one month or one year according to the available data characteristics. From this statistical information a set of indicators will be designed. The Miri@d users can access STAT database to consult the predefined statistics, and they can also access QUERY, DISPLAY and ORDER databases to produce their own statistics.

The QUERY database is constituted by records describing the Article@INIST users' queries. The records of the DISPLAY database contain information about bibliographical references identification and Web user information. The ORDER database is composed by records containing bibliographical references identification and Web user identification. The BIBLIO database contains information about the content of Article@INIST bibliographical references displayed or corresponding to an ordered document. Miri@d users cannot access directly the BIBLIO database. The only function of this database is to supply DISPLAY or ORDER databases with bibliographical data. Figure 1 represents what we call the model. The model is general in the sense that it is not limited to the particular characteristics represented. It is significant to see that the model implies three families of data which it can exploit: on the one hand log-files data and on the other hand bibliographic data and commercial data. From the point of view of scientific information, the bibliographic database can be replaced by any other type of database, at least theoretically. On the level of its concept, the model is more general than as specified by the data sources it is using at the moment.

The server structure consists of a set of external resources providing the statistical raw data, and a set of databases internal to the server. The resources from which Miri@d receives data are the following: Article@INIST server that provides bibliographic database and log-files, document delivery management system, customer management system, and library management system. The Miri@d server own databases are: QUERY data related to Web users queries, DISPLAY data related to displayed bibliographical records, ORDER data related to ordered documents, BIBLIO data related to bibliographical records, STAT data containing calculated statistical indicators on the data stored in the other Miri@d databases. The statistics are calculated off-line with a periodicity of one day, one week, one month or one year.

# 3 New science indicators in the Web context

In comparison with traditional bibliometric studies and in addition to statistics on the bibliographical data, Miri@d is able to treat the requests, i.e. what the users wish to obtain, and how the user expresses his or her request, and the respective top-level domain (TLD, e.g. ".fr" or ".com"), as well as the economic transactions, in which the user becomes customer ordering copies of the retrieved documents. In Miri@d system, bibliometrics is encapsulated inside of web usage analysis, which is the process of applying data analysis techniques to the discovery of usage patterns from Web data. The first step is to create a suitable target data set for the statistical analysis.

## 3.1 Web data source

A Web server log is an important source for performing Web usage analysis and bibliometrics because it explicitly records the browsing behaviour of the site visitors. A set of server sessions is the necessary input for any Web usage analysis. The data recorded in server logs reflects the access of a Web site by multiple users. These log files can be stored in various formats. The Web server stores query data. Query data is generated by online visitors while searching for records (Web pages) relevant to their information needs. The data are obtained from the log-files and in our particular case, the log-files are generated not by the HTTP server, but by the CGI programme interacting with the digital bibliographical database Article@INIST. These log data contain information about both user queries and displayed bibliographic references or records. In these log-files, every request from the user is recorded with different kinds of information depending on the type of action performed.

In a query log-file, we can read for example that on July 20$^{th}$,1999 at 00:51:58AM, a user located in an Australian academic institute, i.e., edu.au, interrogated for articles in English language, published in A and B journals from 1992 to 1999, with an author named Smith. We can also know that 3,306,350 records were been explored in the database and 115, corresponding to the user query, were been retrieved. The user can then choose one or more bibliographic records and ask to display them in an extended version. All this information can now be exploited statistically. (The author affiliation country is not systematically present in Article@INIST bibliographic records. In order to explore this information it will be necessary to complete the records with data coming from external bibliographic databases.)

The information contained in the Article@INIST records is used to obtain bibliometric indicators related to both, records displayed by the user and documents ordered by customers. For Miri@d use only, the information set related to an article is completed with the indication of the scientific domain related to the journal from which the article comes. This data is obtained from LM (Library Management System.) After displaying, the user has the possibility, if he or she is a user-customer, to order directly a copy of the document corresponding to the record. The customer identification permits to link this data with information coming from CM (Customer Management System) in order to obtain customer country, customer sector of activity and other commercial information.

## 3.3 Indicators board

Table 1 shows the sets of statistical data that constitute the output of the Miri@d statistical engine, sets to which analysts can access or interactively produce through the Miri@d server.

| **Web User and Usage** | **Scientific Publication** | **E-commerce** |
|---|---|---|
| Number of queries | Number of displayed records | Number of ordered documents |
| Distribution by:<br><br>- users' TLD<br>- users' country<br>- title word (in query)<br>- author (in query)<br>- keyword (in query)<br>- record displayed | Distribution by:<br><br>- users' TLD<br>- users' country<br>- scientific domain<br>- publication year<br>- author<br>- authors' country<br>- journal<br>- publishing country | Distribution by:<br><br>- customers' country<br>- customers' activity<br>- scientific domain<br>- publication year<br>- author<br>- authors' affiliation country<br>- journal<br>- publishing country |

**Table 1 – Descriptive statistical indicators**

## 3.4 Web usage factors

In addition, we introduce two user indicators dealing with Web users' information retrieval and Web customers' orders. The first is a Web usability factor, and the other is a Web customer order factor. These factors can be considered for evaluating online information sources by the observation of the information displayed by users, and the documents ordered by user-customers. A situation is the number of times an information source is used or displayed by online users. This is a well known situation in information retrieval. The other is the number of times an information source is ordered; in this case we are in face of e-commerce transactions.

### 3.4.1 Web Usability Factor (WUF)

This is the proportion of articles of a journal displayed by Web users in a period of time from $t_0$ to $t_1$ by the total of articles published in this journal and stored until $t_1$.

$$WUF_m = \frac{\sum_{i=t_0,t_1} \sum_{\substack{m=JT \\ \forall n}} dr_i(m,n)}{JT_{t_1}}$$

With: *dr* = displayed records, *JT* = journal title, *n* = publication year, $JT_{t_1}$ = journal title articles, all publication years, stored until $t_1$. The notion of information obsolescence can be also introduced in the calculation of WUF to observe the Web Usability Factor evolution by publication year.

$$WUF(PY)_m = \frac{\sum_{i=t_0,t_1} \sum_{\substack{m=JT \\ n=PY}} dr_i(m,n)}{JT_{t_1}(PY)}$$

With: $dr$ = displayed records, $[t_0, t_1]$ = period of time, $JT$ = journal title, $PY$ = publication year considered, $JT_{t_1}$ = number of journal title articles in publication year PY, stored until $t_1$.

### 3.4.2 Customer Order Factor (COF)

This is the proportion of articles of a journal ordered by Web customers in a period of time from $t_0$ to $t_1$ by the total number of articles published in this journal and stored until $t_1$.

$$COF_m = \frac{\sum_{i=t_0,t_1} \sum_{\substack{m=JT \\ \forall n}} ord_i(m,n)}{JT_{t_1}}$$

With: $ord$ = ordered documents, $JT$ = journal title, $N$ = publication year, $JT_{t_1}$ = journal title articles, all publication years, stored until $t_1$. We can also introduce the notion of information obsolescence in the calculation of COF to observe the Customer Order Factor evolution by publication year.

$$COF(PY)_m = \frac{\sum_{i=t_0,t_1} \sum_{\substack{m=JT \\ n=PY}} ord_i(m,n)}{JT_{t_1}(PY)}$$

With: $ord$ = ordered documents, $[t_0, t_1]$ = period of time, $JT$ = journal title, $PY$ = publication year considered, $JT_{t_1}$ = number of journal title articles in publication year PY, stored until $t_1$.

# 4 Application of Miri@d

This section deals with the presentation of Miri@d double application: on the one hand the statistical database placed at the disposal of users and on the other hand, the interactive application with which users can carry out their own statistical analyses on the bibliographical data, on the requests of the Web users and on economic data related to document orders. The off-line predefined statistical application is presented in section 4.1.1 and section 4.1.2 proposes an example of interactive application.

## 4.1 Statistics database

The Miri@d server proposes predefined statistics calculations stored in the STAT database, where users can consult statistical information updated regularly. The information exploited in this STAT database comes from the QUERY, DISPLAY and ORDER databases and corresponds respectively to web users' searches data, target bibliographic data and economic data. The Miri@d predefined statistics are grouped into these three sub-sets.

We will present as example the distribution of web users' searches by users' country during the year 2002 (table 2). There are 135 different users' countries with a very large predominance of users from France, more than 77% of searches. About 87.5% of searches come from the European Union, the USA and Japan, respectively 84.4%, 2.3% and 0.8%. The remaining 12.5% of searches are shared by the other 118 countries. The 15 countries constituting the European Union are at the origin of 872,510 searches corresponding to more than 84%, and around 92% of these searches come from France. The 8% of searches, corresponding to 73,375 queries, show Spain in the first rank with 16,652 queries (22.7%) and Ireland in the last one with only 573 queries (0.8%).

| Rank | Web user country | Number of queries | Percentage of queries |
|---|---|---|---|
| 1 | France | 799,135 | 77.31 |
| 2 | Canada | 33,414 | 3.23 |
| 3 | United States | 23,461 | 2.27 |
| 4 | Morocco | 23,353 | 2.26 |
| 5 | Spain | 16,652 | 1.61 |
| 6 | Belgium | 16,630 | 1.61 |
| 7 | Germany | 12,052 | 1.17 |
| 8 | Tunisia | 11,262 | 1.09 |
| 9 | Switzerland | 11,079 | 1.07 |
| 10 | Italy | 10,845 | 1.05 |
| … | … | … | … |

**Table 2 - Number of web users' searches by country**

Concerning displayed records, the following statistics are available: the number of records by document type, the N most often displayed articles, and the N most often displayed journals (the value of N depends on the periodicity selected by Miri@d users). Table 3 presents the 10 most often displayed journals for the year 2002 as well as their WUF.

| Rank | Number of visualisations | Journal title | Web Usability Factor |
|---|---|---|---|
| 1 | 3086 | Macromolecules | 0.16 |
| 2 | 2858 | Journal of applied polymer science | 0.18 |
| 3 | 2546 | Polymer | 0.21 |
| 4 | 2495 | Langmuir | 0.16 |
| 5 | 1798 | Journal of applied physics | 0.05 |
| 6 | 1529 | Physical review letters | 0.04 |
| 7 | 1470 | Applied physics letters | 0.04 |
| 8 | 1436 | Physical review B, Condensed matter & materials physics | 0.05 |
| 9 | 1393 | La Presse médicale | 0.19 |
| 10 | 1212 | Journal of colloid and interface science | 0.14 |
| … | … | … | |

**Table 3 - Number of visualisations by displayed journals**

Concerning ordered documents (see Table 1) the following statistics are available: the number of ordered documents by customers' country, the number of ordered documents by customers' type of activity, and also the 50 most often ordered articles and ordered journals. Table 4 presents the distribution of ordered documents by type of customers' activity for the year 2002. The typology of users' activities adopted by Miri@d is the one used by INIST's Customer Management System.

| Rank | Web customer activity domain | Number of ordered documents | Percentage |
|---|---|---|---|
| 1 | Commercial firms | 38,333 | 53.8 |
| 2 | Research institutions | 20,070 | 28.2 |
| 3 | Higher education | 8,996 | 12.6 |
| 4 | Others institutions | 1,625 | 2.3 |
| 5 | Hospitals | 769 | 1.1 |
| 6 | Information centers | 736 | 1.0 |
| 7 | Private person | 679 | 0.9 |
| | Total | 71,208 | 100.0 |

**Table 4 - Number of ordered documents by Web customers' activity**

Table 5 shows the 10 first most often ordered journals in the considered period (2002) as well as their COF. This statistical information allows us to have an image of the customers' interests.

| Rank | Number of orders | Journal title | Customer Order Factor |
|---|---|---|---|
| 1 | 366 | Journal of applied polymer science | 0.022 |
| 2 | 228 | Journal of agricultural and food chemistry | 0.019 |
| 3 | 221 | Journal of chromatography | 0.015 |
| 4 | 202 | Langmuir | 0.013 |
| 5 | 183 | Polymer | 0.015 |
| 6 | 178 | Lancet : (British edition) | 0.007 |
| 7 | 176 | La Revue du praticien | 0.029 |
| 8 | 168 | Journal of applied physics | 0.004 |
| 9 | 167 | Journal of materials science | 0.013 |
| 10 | 163 | Journal of the American Chemical Society | 0.005 |
| … | … | … | … |

**Table 5 - Number of orders by ordered journals**

## 4.2 Obtaining your own statistics

In order to produce their own statistical analysis, Miri@d users can access the QUERY, DISPLAY and ORDER databases containing respectively web users' searches data, target bibliographic data and economic data.

The statistical analysis of keywords employed by users in their queries gives information about the subjects interesting users. To illustrate the QUERY database utilization, we selected a set of queries about the word polymer in a journal title. The number of queries by users' country is shown in Table 6.

| User's country | Number of queries | Percentage of queries |
|---|---|---|
| France | 2,061 | 82.1 |
| Iran | 63 | 2.5 |
| Tunisia | 61 | 2.4 |
| Belgium | 42 | 1.7 |
| Germany | 42 | 1.7 |
| Morocco | 28 | 1.1 |
| United States | 27 | 1.1 |
| Australia | 26 | 1.0 |
| Brazil | 23 | 0.9 |
| Spain | 21 | 0.8 |
| Others (18 countries) | 117 | 4.6 |

**Table 6 - Number of web users' searches by country**

The interrogation of the DISPLAY database allows Miri@d users to access the information about the records displayed by the web users of the Article@INIST repository. To illustrate the DISPLAY database utilization, we use a set of 917 records displayed by users and linked to the above mentioned set of queries about polymers. Table 7 shows the distribution of these records by users' countries.

| User's country | Number of displayed records | Percentage of displayed records |
|---|---|---|
| France | 726 | 79.2 |
| Belgium | 115 | 12.5 |
| Canada | 17 | 1.9 |
| Brazil | 16 | 1.7 |
| Venezuela | 7 | 0.8 |
| Germany | 6 | 0.7 |
| Morocco | 5 | 0.5 |
| Others (11 countries) | 25 | 2.7 |

**Table 7 - Number of displayed records by users' countries**

The country with the greatest number of displayed records is France with 79% of the total. Seven other countries belonging to the European Union are represented, particularly Belgium with 115 records' visualisations corresponding to 12%. The total number of displayed journals is equal to 82 and Table 8 presents the 10 most often displayed journals as well as their WUF for the year 2002.

| Journal | Number of visualisations | Percentage of visualisations | Web Usability Factor |
|---|---|---|---|
| Journal of applied polymer science | 113 | 12.3 | 0.18 |
| Polymer | 77 | 8.4 | 0.21 |
| Polymer engineering & science | 52 | 5.7 | 0.20 |
| European polymer journal | 35 | 3.8 | 0.18 |
| Colloid & polymer science | 33 | 3.6 | 0.24 |
| Journal of polymer science. Part A | 31 | 3.4 | 0.13 |
| Journal of polymer science | 30 | 3.3 | 0.18 |
| Polymer preprints, American Chemical Society | 30 | 3.3 | 0.21 |
| Advances in polymer science | 27 | 2.9 | 0.19 |
| Polymer degradation and stability | 23 | 2.5 | 0.20 |
| Others (72 journals) | 466 | 50.8 | |

**Table 8 - Number of visualisations by displayed journal**

The interrogation of the ORDER database allows Miri@d users to access a database collecting the information about both the customers ordering the documents and the documents themselves. The distribution of the 4,513 ordered documents by customers' country is presented in Table 9, and the distribution of the 4,513 ordered documents by customers' activity is given in Table 10.

| Customer's country | Number of ordered documents | Percentage of ordered documents |
|---|---|---|
| France | 3,878 | 85.9 |
| Belgium | 511 | 11.3 |
| Brazil | 79 | 1.7 |
| Russia | 15 | 0.3 |
| Germany | 9 | 0.2 |
| Monaco | 6 | 0.1 |
| Others (9 countries) | 15 | 0.3 |

**Table 9 - Number of ordered documents by customers' country**

| Rank | Customer's activity | Number of ordered documents | Percentage of ordered documents |
|---|---|---|---|
| 1 | Commercial firm | 2,072 | 45.9 |
| 2 | Research institutions | 2,033 | 45.0 |
| 3 | Higher education | 373 | 8.1 |
| 4 | Others institutions | 301 | 0.5 |
| 5 | Information centers | 25 | 0.3 |
| 6 | Private person | 24 | 0.1 |
| | Total | 4.513 | 100.00 |

**Table 10 - Number of ordered documents by customers' activity**

The most represented customer activity is "Commercial firm" with 46% of orders, but "Research institutions" is not far behind with 45%. The only other significant value belongs to "Higher education" which represents 8%.

# 5 Co-usage analysis

Co-usage analysis belongs to the same family that co-citation and co-word analysis. We expose in this section the formal characteristics of co-usage analysis, which obeys the co-occurrence framework for clustering and mapping the Web usage. We use it for identifying and visualising usage centres of interest on certain research foci, or problem areas. Co-usage analysis will be applied in the next section on the Web users' data produced by the Miri@d server. Actually, co-usage analysis is not yet included in Miri@d. Another Web server we called VISA, makes it possible for analysts to reach the co-usage analysis results.

On the basis of vector-space model, co-usage analysis consists in producing, from the Web users' data provided by Miri@d, a normalised co-occurrence matrix, i.e. an association matrix. For this purpose, an association coefficient will be used. The next step is to apply a clustering method and to draw a map. In the following sub-sections we detail this approach.

## 5.1 Vector-Space Model

A mathematical model, to represent either users or visualised records during a query session, is the well known *vector-space model* (Salton, 1989, chapter 10). The vector-space model assumes that an available attribute set is used to identify both stored records and information requests. Both queries and documents can then be represented as attribute vectors. For a particular document, it is possible to identify a set of user references by log data analysis, as well as a set of references made by other users to the document in question. When user $u$ refers in his or her query to bibliographic data $d$ implies that document $d$ is included as an entry in the usage of user $u$.

## 5.2 Co-usage concept

Documents $d_i$ and $d_j$ are related by usage coupling, when a user $u_i$ refers in his or her query to documents $d_i$ and $d_j$. If $u_i \rightarrow d_i$ and $u_i \rightarrow d_j$, then $d_i$ and $d_j$ are associated $d_{ij}$ by $u_i$, with $i$ = 1, 2, …, $m$. Conversely, users $u_i$ and $u_j$ are related as a pair of users, $u_{ij}$, when they refer in their queries to a same document $d_i$. If $u_i \rightarrow d_i$ and $u_j \rightarrow d_i$, then $u_i$ and $u_j$ are associated by $d_i$, with $i$ = 1, 2, …, $n$. A given document or bibliographic data is represented by an usage attribute vector of the form: $d_i = (u_{i1}, u_{i2}, \ldots, u_{im})$, where $u_{ij}$ is an identifier representing the *jth* user reference of the document $i$. Vector similarity operation is performed using user references similarities. A given user can be represented by a record (bibliographic data) attribute vector of the form: $u_i = (d_{i1}, d_{i2}, \ldots, d_{in})$, where $d_{ij}$ is an identifier representing the *jth* document referred by user $i$ and displayed or ordered by the same the user $i$. Vector similarity can be performed using displayed documents similarities.

Thus, each document $d_i$ can be described by the set of co-users $u_{ij}$, $d_i = \left[u_{ij}\right]_{i=1,n;\, j=1,m}$; with $d_{ij} \in [0,1]$; and $m$ = number of users, $n$ = number of documents. Then the document co-occurrence matrix denoted by $D$ is:

$$D = \begin{pmatrix} u_{11} & \cdots & u_{1m} \\ \vdots & \ddots & \vdots \\ u_{n1} & \cdots & u_{nm} \end{pmatrix}$$

In the same way, each user can be described by a set of co-documents $d_{ij}$, $u_i = \left[d_{ij}\right]_{i=1,m;\, j=1,n}$; with $d_{ij} \in [0,1]$; and $m$ = number of users, $n$ = number of documents. Then the user co-occurrence matrix denoted by $U$ is:

$$U = \begin{pmatrix} d_{11} & \cdots & d_{1n} \\ \vdots & \ddots & \vdots \\ d_{m1} & \cdots & d_{mn} \end{pmatrix}$$

An association coefficient is used for normalising the co-occurrences in $D$ and $U$ matrices. There are several methods of calculating an association coefficient. We use the so called "equivalence coefficient" largely used in co-word analysis and which it is defined as:

$$E_{ij} = \frac{C_{(i,j)}^{\ 2}}{o_{(i)} \times o_{(j)}}$$

*C(i,j)* is the total number of co-occurrences of users *i* and *j* or documents *i* and *j*, *o(i)* is the total number of occurrences of item *i*, *o*(*j*) is the total number of occurrences of user *b*. This association coefficient is analogous to the well known Dice, Jaccard, Ochiai, or Salton coefficients. As results of the application of this association coefficient on user couples, we obtain a real valued matrix, denoted by *A*, and composed of the association coefficients values. The association matrix *A* gives a normalized measure of the strength of associations between users *AU*, and between documents *AD*. We apply a single linkage hierarchical agglomerative clustering method to these matrices.

The algorithm we use is an adaptation of the standard bottom-up single-link clustering in accordance with readability criteria on the size of the cluster, which is defined as the minimum and maximum number of items belonging to the cluster, and on the maximum number of associations constructing the cluster. As a consequence of this clustering process, two kinds of associations and items forming the clusters are generated: one is said to be internal or intra-cluster and the other external or inter-cluster. If both elements of a given pair belong to the same cluster, the association between the items is considered as an internal association of that cluster. If they belong to two different clusters, the association is considered as an external association. The items involved in internal associations of a cluster are called internal items. The number of internal items defines the size of a cluster. Those items rejected during clustering because they do not meet the “maximum cluster size” criterion are recorded as external items.

## 5.3 Indicator measures

Each clustered item presents a weight indicating its prominence in the cluster. Let be *a* an internal or external item (document or user) of the cluster *Cl*, then the weight is defined as follows:

$$\underset{a=1,m_{Cl}}{w_{Cl}}(a) = \frac{k_{Cl\,(a)}}{n_{Clin} + n_{Clex}} \text{ with: } 0 < k_{Cl}(a) \le n_{Clin} + n_{Clex} \text{ And: } 0 < w_{Cl}(a) \le 1$$

Symbol sense: $m_{Cl}$ is the number of its internal and external items, $n_{Clin}$ is the number of its internal associations, $n_{Clex}$ is the number of its external associations, $k_{Cl}(a)$ is the number of occurrences of internal or external item *a* $(a = 1, m_{Cl})$ in the internal or external associations of *Cl.*

A relevance weight $r_{Cl}(i)$) is also computed for each information source unit related to the considered cluster. This is defined as the sum of the item weights $w_{Cl}(a)$ of the items belonging to both, the set of items present in the information source unit, and the internal items cluster set, divided by the total number of items present in the information source unit. It is possible to observe some information source units with multiple values of relevance weight. Indeed, as the same information source unit contributes often to two and plus clusters, it can have a different calculated relevance weight in each cluster. Let *Cl* be a cluster and $m_{Clin}$ = the number of its internal items; $l_{Cl}(i)$ = the number of its internal items present in the source information unit *i*; $s_{Cl}$ = the number of source information units contributing to the cluster *Cl*; *L(i)* = the number of items present in the source information unit *i*. Then, for each source information unit *i* related to *Cl* we can calculate its relevance weight *r* as:

$$\underset{i=1,s_{Cl}}{r_{Cl}}(i) = \frac{\sum_{a=1,l_{Cl}(i)} w_{Cl}(a)}{L(i)} \text{ with: } 0 < l_{Cl}(i) \le m_{Clin} \ ; \ 0 < w_{Cl}(a) \le 1 \text{, and: } 0 < r_{Cl}(i) \le \frac{m_{Clin}}{L(i)}$$

In addition, the clusters are characterized by two structural properties respectively called *density* and *centrality*. Cluster density $D_{Cl}$ is defined as the mean value of the internal associations (intra-cluster). The density is an indicator of the cohesiveness of the clusters. Cluster centrality $C_{Cl}$ is defined as the mean value of the external associations (inter-clusters). The centrality is an indicator of the position of clusters in the network of inter-cluster relationships. Note that these notions of density and centrality are a little different from those usually employed in graph theory. A structural measure $S_{Cl}$ can be also calculated as the ratio of centrality to density. A low value of this indicator often indicates that a cluster can be cohesive but with a weak centrality. A strong value of this indicator often indicates that a cluster can be located as central and thus to be strategic but with little cohesion (Courtial 1990). Centrality and density have been largely used in co-word analysis. We extended them into our co-usage analysis made possible now by the existence of Web log-file data.

On the other hand, the measures of *density* and *centrality* allow the visualization of clusters and their relationships in a two-dimensional space: a map, where the X-axis corresponds to *centrality* and the Y-axis to *density*. The map allows us to understand the global and the local structure brought out by the clustering method from the association matrices.

# 6 Co-usage analysis application

We have insisted above on the fact that Miri@d provides descriptive statistics information that we can use to carry out more advanced tasks of analysis. We apply the co-usage analysis (which was described in section 4) to data coming from the Miri@d server. A detailed interpretation of the results is out of question in this article. We limit here to expose the results that the co-usage analysis provides to analysts who would perform the interpretation task. However, we will underline especially the means of analysis provided to assist the work of interpretation and decision.

From Miri@d's ORDER database, a set of documents dealing with polymers was collected. These data consist of scientific articles that have been ordered by customers after consultation of the Article@INIST repository. The size of the ORDER database is the 202,391 orders through two years between 2001 and 2003. For this application, the dataset is equal to 3,914 documents published by 57 journals and ordered by 410 customers.

Since each document $d_i$ can be described by the set of customers $u_{ij}$, and reversely each user $u_i$ can be described by the set of documents $d_{ij}$. Two co-usage analyses have been done: one on the ordered documents and the other on the user-customers (or authors of the orders.) In the next sections, we describe these applications at the levels of clusters (see figures 3 and 4) and maps (see figures 5 and 6).

## 6.1 Co-usage analysis of the ordered documents

24 clusters were obtained from the ordered-document co-occurrence matrix $U(\mathrm{d}_{ij})$ $i = 1,\ldots,m$ and $j = 1,\ldots,n$; $m$ = number of user-costumers = 410, $n$ = number of ordered-documents = 3914. They are constituted by ordered-documents dealing with subjects interesting sub-sets of users. The documents are represented by codes. Figure 3 shows the graph structure of a cluster. The map is represented in figure 5.

## 6.2 Co-usage analyse of the user-customers

17 clusters were obtained from the user-costumers co-occurrence matrix $D(u_{ij})$ $i = 1,\ldots,n$ and $j = 1,\ldots,m$; $m$ = number of user-customers = 410, $n$ = number of ordered-documents = 3914. These clusters are then constituted by user-customers who are interested by sub-sets of ordered-documents in which the knowledge interesting these user-customers is embedding. Figure 4 shows an example of the graph of a user-customer cluster. To preserve confidentiality the customers are represented only by their country and activity sector codes. The map of the user-customer clusters is presented in figure 6.

## 6.3 Analytical tools: clusters and maps

Clusters and maps constitute analytical tools. A cluster is composed of items that are called internal items. The internal item with the maximal weight value $w_{Cl}(a)$ is automatically chosen to be the cluster label. The clusters are also composed of associations between these items which are also called internal associations, to distinguish them from external associations which link a cluster with other clusters.

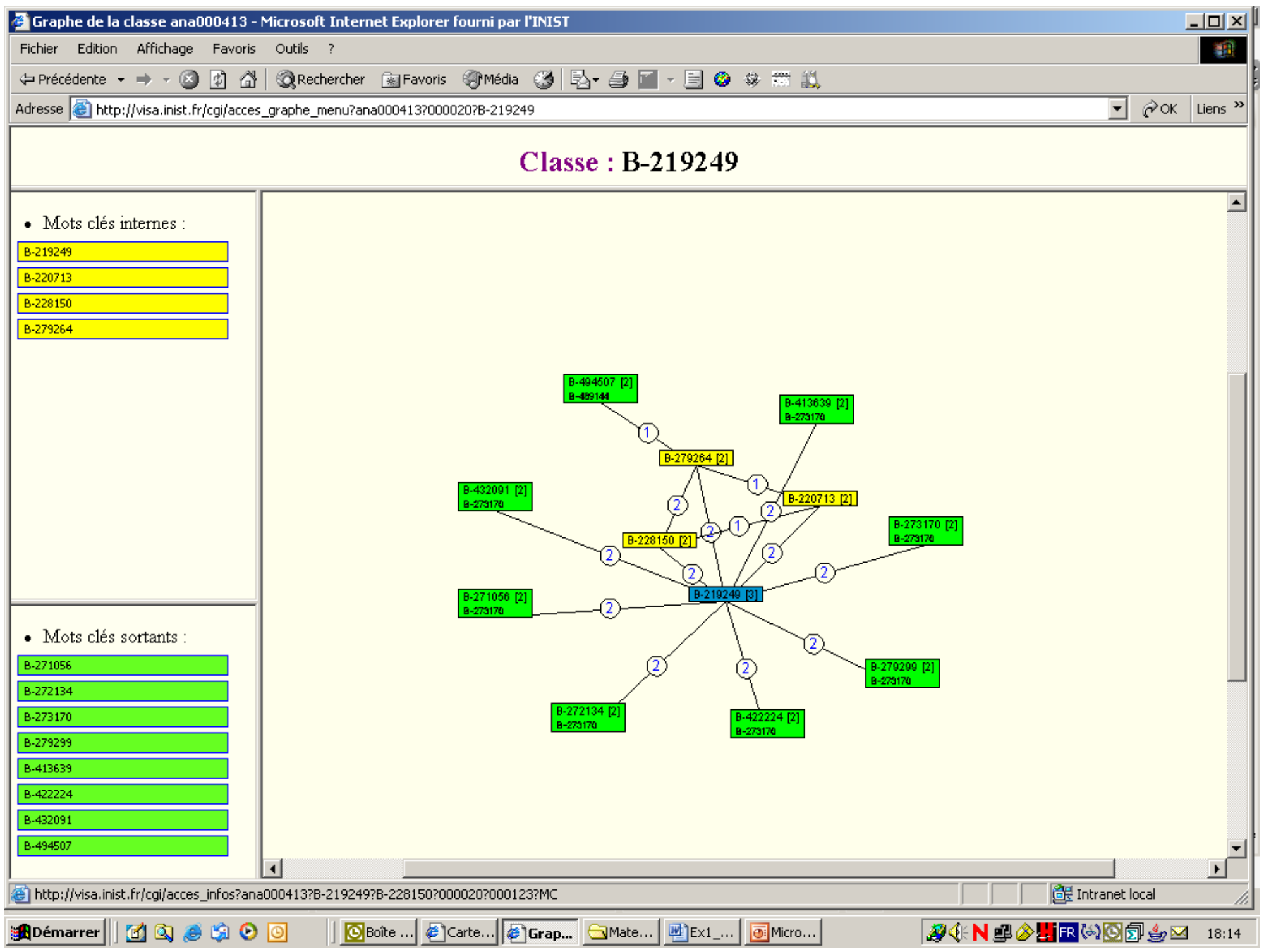

Figure 3: Cluster graph labelled by B-219249 ordered document

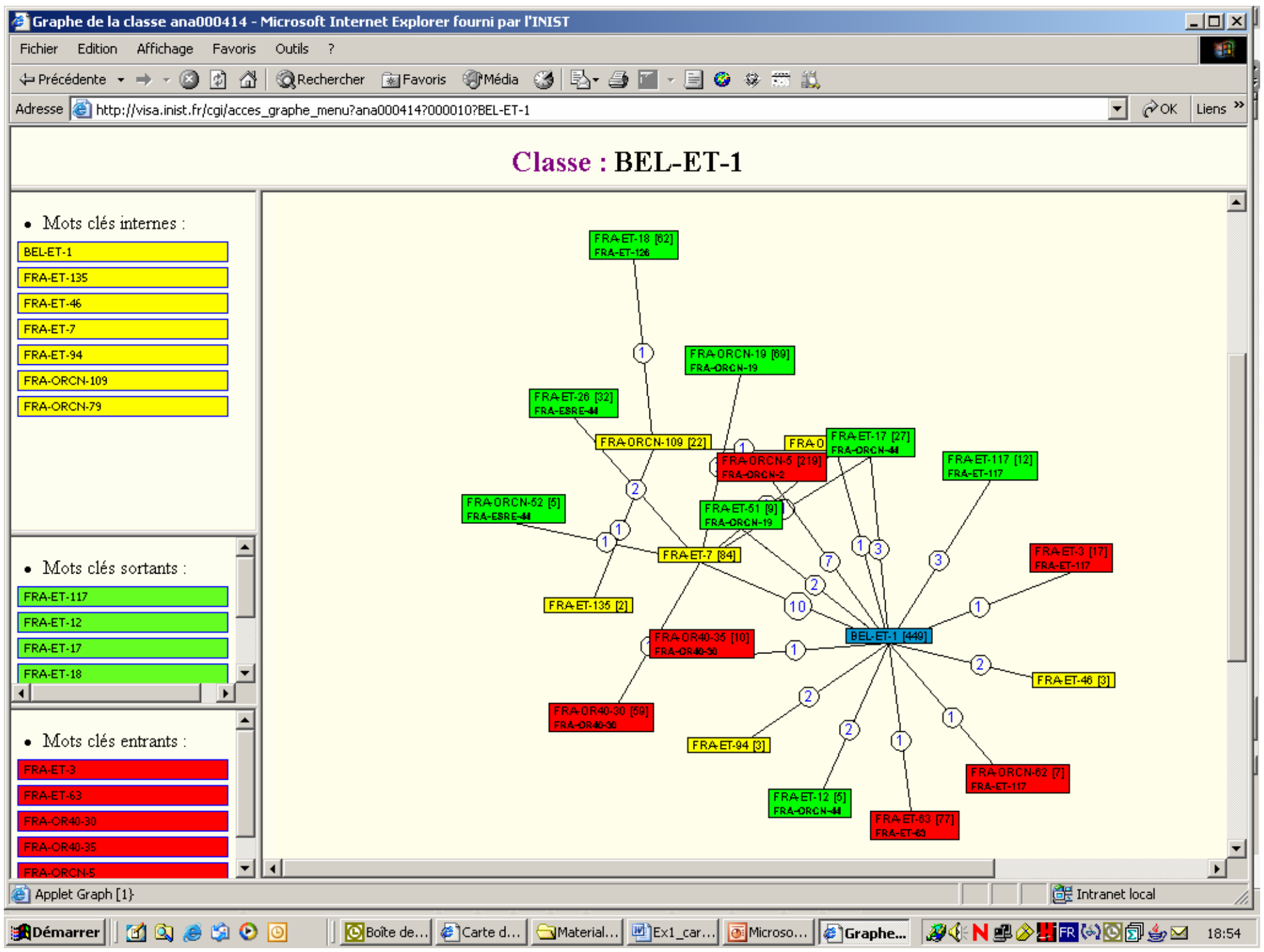

Figure 4: Cluster graph labelled by BEL-ET-1 user-customer

The internal and external associations are weighted relations according to $E_{ij}$ association coefficient. The internal associations between the component items of a cluster are represented by the graphs in figures 3 and 4. On the other hand, the links between clusters are shown on the maps of the figures 5 and 6.

In figures 3 and 4 two classes of internal items are defined: leaving and entering. This allows us to know by which item this cluster is connected to another cluster, and at the same time which are the internal items by which the cluster receives links starting in another cluster. According to this angle of analysis, the clusters can be analysed as insulated, receivers or senders within the network of clusters.

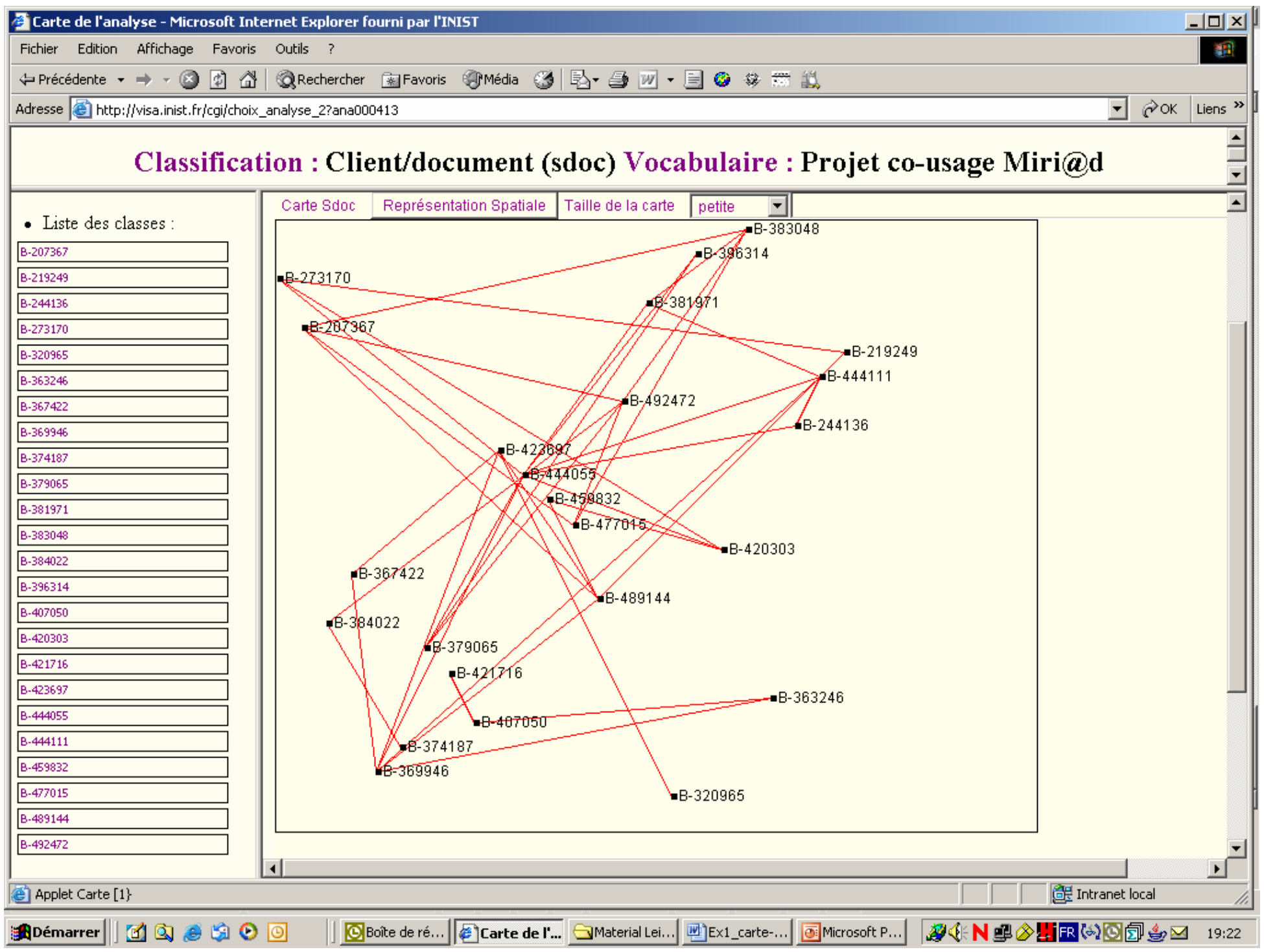

Figure 5: Map 1 - Ordered document clusters

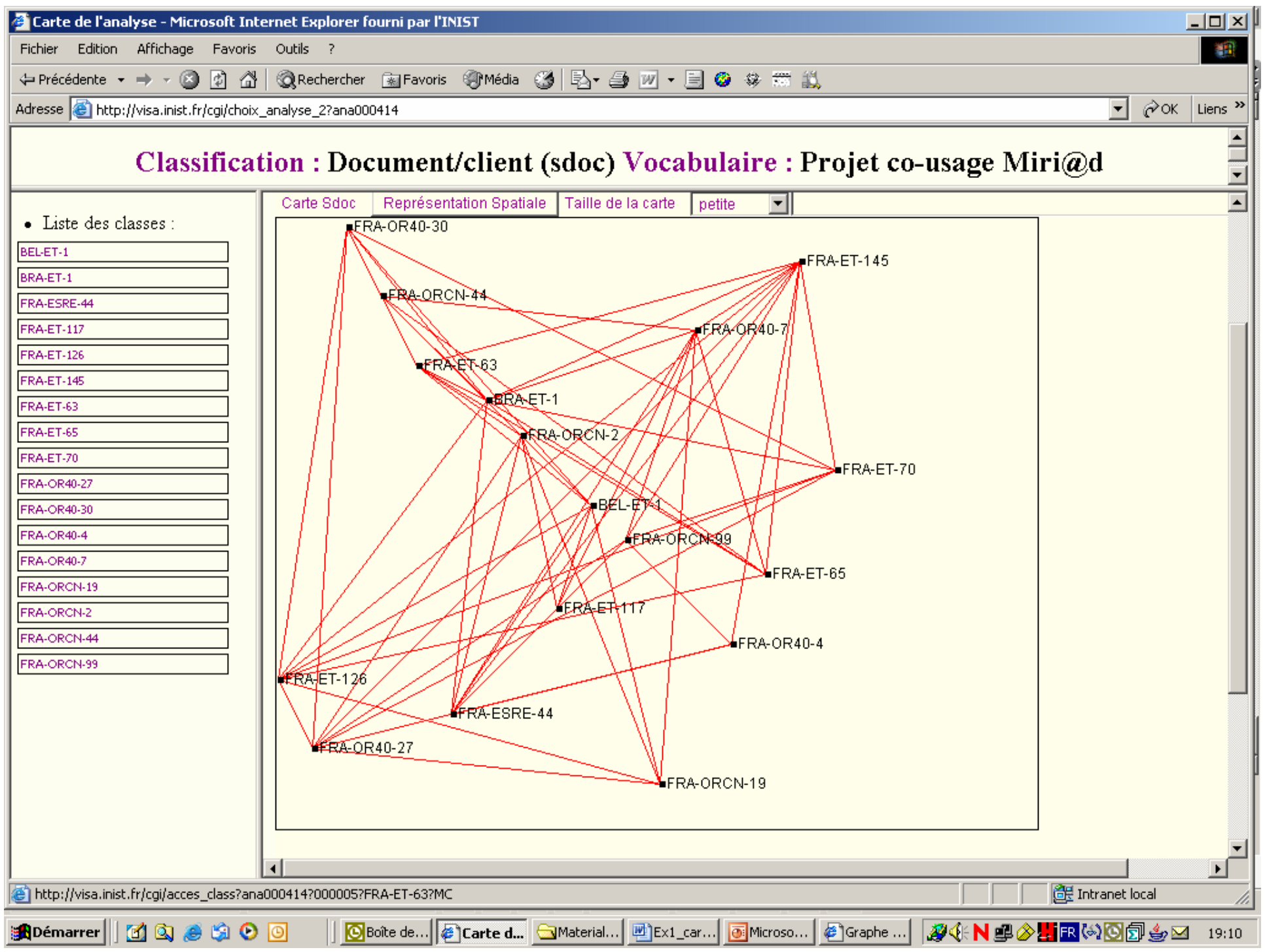

Figure 6: Map 2 - User-customer clusters

On the maps, clusters are placed in function of their structural properties, that is centrality (X-axis) and density (Y-axis). Each labelled point in the maps represents a cluster. Following the standard co-word analysis, four types of clusters can be distinguished: clusters with high density and centrality (type 1), with low density and high centrality (type 2), with high density and low centrality (type 3), and clusters with low values on both axes (type 4). The map (called strategic diagram) is then analysed in these terms.

On the other hand, clusters appear within a network. We can visualise in the maps the network of inter-cluster relations. Thus, the co-usage analysis can be extended as a network analysis using graph theory, and then we can interpret clusters and their network in this other analytical framework (Polanco 2005). What is significant here is to see that we have two possibilities of analysis, one founded in the tradition of the representation of the classes in a two-dimensional plan, the other according to the network analysis and the theory of graphs.

# 7 Conclusion

We presented in this article the Miri@d server, which is conceived to produce descriptive statistical data about what is effectively used (in a free access digital database storing scientific and technical information), and the Web users searching behaviour. We showed that the system is conceived as a server of statistical data which are carried out beforehand, and as an interactive server for personal online statistical work. Two Web usage factors were also proposed.

We underlined that Miri@d represents a model that is not limited to the particular characteristics of the digital database to which it is connected. We drew the attention on the fact that the model implies three families of data: log-files data, bibliographic data (which are equivalent to products), and commercial data about documents orders.

Over all, we focused in this article to define and to discuss the instrument with its indicator board, as well as the aspects of method about the analysis of the users. At the same time, we tried to show how today with the Web new ways emerging for analysing science through its publications and their usages. It is the existence of log-files that provides the technical opportunity for introducing the analysis of user behaviours and proposing then the co-usage analysis.

We consider the co-usage analysis as a generalisation of the co-word analysis, and we remark that the same approach can be used at the level of both the Web content analysis and the Web structure analysis (Polanco et al 2001).

**Acknowledgment:** The design and development of the MIRI@D server, as well as the indicators engineering, and also the co-usage analysis proposition, have been realized thanks to the European project IST-1999-20350 - Fifth Research and Development Framework Plan of the European Union, during 2000-2003. Project acronym EICSTES, and project full title “European Indicators, Cyberspace, and the Science – Technology – Economy System.”

# 8 Bibliographie

[1] ALMIND T.C. & INGWERSEN P., *Informetric analyses on the World Wide Web: methodological approaches to “webometrics”*, Journal of Documentation, vol. 53, No 4, p. 404-426, 1997

[2] BJÖRNEBORN L. & INGWERSEN P., *Perspectives of webometrics*, Scientometrics, vol. 50, No 1, p. 65-82, 2001

[3] CALLON M. and COURTIAL J.-P., *Using scientometrics for evaluation*, in M. Callon, Ph. Larédo, Ph. Mustar (editors), The Strategic Management of Research and Technology, Paris, Economica International, chapter 10, p. 165-219

[4] CHAKRABARTI S., *Mining the Web*, Morgan Kaufmann Publishers, 2003

[5] COURTIAL J.-P., *Introduction à la scientométrie*, Paris, Anthropos, 1990

[6] INGWERSEN P. & BJÖRNEBORN L., *Methodological issues of webometric studies*, in Handbook of Quantitative Science and Technology Research. Edited by H.F. Moed, W. Glänzel and U. Schmoch, Kluwer Academic Publishers, chapter 15, p. 339-369, 2004

[7] KOSALA R. & BLOCKEEL H., *Web Mining Research: A Survey*, SIGKDD Explorations, vol. 2, No 1 p. 1-15, 2000 (available at: http://portal.acm.org/)

[8] POLANCO X., *Modelling co-word clusters as graphs*, 2005 (unpublished preprint)

[9] POLANCO X., BOUDOURIDES M. A., BESAGNI D., ROCHE I., *Clustering and Mapping European University Web Sites Sample for Displaying Associations and Visualizing Networks*, in Pre-proceedins New Techniques and Technologies for Statistics (NTTS&ETK), 18-22 June, Crete, vol. 2 p. 941-944, 2001

[10] SALTON G., *Automatic Text Processing*, Reading, Mass., Addison-Wesley Publishing, 1989